# THE SINS/zC-SINF SURVEY OF z ~ 2 GALAXY KINEMATICS: EVIDENCE FOR POWERFUL AGN-DRIVEN NUCLEAR OUTFLOWS IN MASSIVE STAR-FORMING GALAXIES[*]

N.M. Förster Schreiber[1], R. Genzel[1,2,3], S.F. Newman[3], J.D. Kurk[1], D. Lutz[1], L.J. Tacconi[1], S. Wuyts[1], K. Bandara[1], A. Burkert[4], P. Buschkamp[1], C.M. Carollo[5], G. Cresci[6], E. Daddi[7], R. Davies[1], F. Eisenhauer[1], E.K.S. Hicks[8,9], P. Lang[1], S.J. Lilly[5], V. Mainieri[10], C. Mancini[11], T. Naab[12], Y. Peng[13], A. Renzini[11], D. Rosario[1], K. Shapiro Griffin[14], A.E. Shapley[15], A. Sternberg[16], S. Tacchella[5], D. Vergani[17], E. Wisnioski[1], E. Wuyts[1] & G. Zamorani[6]

[1] Max-Planck-Institut für extraterrestrische Physik, Giessenbachstrasse, D-85748 Garching, Germany
[2] Department of Physics, Le Conte Hall, University of California, Berkeley, CA 94720, USA
[3] Department of Astronomy, Hearst Field Annex, University of California, Berkeley, CA 94720, USA
[4] Universitäts-Sternwarte, Ludwig-Maximilians-Universität, Scheinerstrasse 1, D-81679 München, Germany
[5] Institute for Astronomy, Department of Physics, Eidgenössische Technische Hochschule, CH-8093 Zürich, Switzerland
[6] Istituto Nazionale di Astrofisica – Osservatorio Astronomico di Bologna, via Ranzani 1, I-40127 Bologna, Italy
[7] CEA Saclay, DSM/IRFU/SAp, F-91191 Gif-sur-Yvette, France
[8] Department of Astronomy, University of Washington, Box 351580, U.W., Seattle, WA 98195-1580, USA
[9] Department of Physics and Astronomy, University of Alaska Anchorage, 3211 Providence Drive, Anchorage, AK 99507, USA
[10] European Southern Observatory, Karl-Schwarzschild-Strasse 2, D-85748 Garching, Germany
[11] Istituto Nazionale di Astrofisica – Osservatorio Astronomico di Padova, Vicolo dell'Osservatorio 5, I-35122 Padova, Italy
[12] Max-Planck-Institut für Astrophysik, Karl-Schwarzschild-Strasse 1, D-85748 Garching, Germany
[13] Astrophysics Group, Cavendish Laboratory, JJ Thomson Avenue, Cambridge, CB3 0HE, United Kingdom
[14] Space Sciences Research Group, Northrop Grumman Aerospace Systems, Redondo Beach, CA 90278, USA
[15] Department of Physics and Astronomy, University of California, Los Angeles, CA 90095-1547, USA
[16] School of Physics and Astronomy, Tel Aviv University, Tel Aviv 69978, Israel
[17] Istituto Nazionale di Astrofisica – Istituto di Astrofisica Spaziale e Fisica Cosmica, Via Gobetti 101, I-40129 Bologna, Italy

## ABSTRACT

We report the detection of ubiquitous powerful nuclear outflows in massive ($\geq 10^{11}$ $M_{\odot}$) z ~ 2 star-forming galaxies (SFGs), which are plausibly driven by an Active Galactic Nucleus (AGN). The sample consists of the eight most massive SFGs from our SINS/zC-SINF survey of galaxy kinematics with the imaging spectrometer SINFONI, six of which have sensitive high-resolution adaptive optics (AO) assisted observations. All of the objects are disks hosting a significant stellar bulge. The spectra in their central regions exhibit a broad component in H$\alpha$ and forbidden [NII] and [SII] line emission, with typical velocity FWHM ~ 1500 km s$^{-1}$, [NII]/H$\alpha$ ratio $\approx$ 0.6, and intrinsic extent of 2 – 3 kpc. These properties are consistent with warm ionized gas outflows associated with Type 2 AGN, the presence of which is confirmed via independent diagnostics in half the galaxies. The data imply a median ionized gas mass outflow rate of ~ 60 $M_{\odot}$ yr$^{-1}$ and mass loading of ~ 3. At larger radii, a weaker broad component is detected but with lower FWHM ~ 485 km s$^{-1}$ and [NII]/H$\alpha$ $\approx$ 0.35, characteristic for star formation-driven outflows as found in the lower-mass SINS/zC-SINF galaxies. The high inferred mass outflow rates and frequent occurrence suggest the nuclear outflows efficiently expel gas out of the centers of the galaxies with high duty cycles, and may thus contribute to the process of star formation quenching in massive galaxies. Larger samples at high masses will be crucial to confirm the importance and energetics of the nuclear outflow phenomenon, and its connection to AGN activity and bulge growth.

---

## 1. INTRODUCTION

Feedback is thought to play a critical role in the evolution of galaxies by regulating their stellar mass and size growth, modulating their global star formation and chemical enrichment history, and mediating the relationship between the growth of supermassive black holes and their host galaxies (e.g., Kauffmann et al. 1993; Silk & Rees 1998; Efstathiou 2000; Croton et al. 2006; Sales et al. 2010; Davé et al. 2012; Hopkins et al. 2012; Lilly et al. 2013; Vogelsberger et al. 2013; Hirschmann et al. 2013). Feedback is generally acknowledged as a critical ingredient to explain the low baryon conversion efficiency of galaxies at low and high masses (e.g., Baldry et al. 2008; Moster et al. 2010; 2013) but the actual physical mechanisms involved and the relative importance of star formation versus AGN as main drivers are still poorly known. At the high mass end, and out to redshifts $z \approx 4$, the abundance of galaxies above a characteristic stellar mass of $\log(M_*/M_\odot) \sim 10.8$ drops exponentially and star formation seems to be efficiently shut off in a significant proportion of galaxies, which rapidly become quiescent (e.g., Peng et al. 2010; Brammer et al. 2011; Ilbert et al. 2013). Feedback from active galactic nuclei (AGN) has been widely invoked as an agent for this "quenching" (e.g., Di Matteo et al. 2005; Croton et al. 2006; Bower et al. 2006; Somerville et al. 2008) but little observational evidence exists on how such feedback operates at high redshift (see the review by Fabian 2012).

Observationally, the widespread detection of blueshifted rest-UV ionic absorption lines demonstrates that most star-forming galaxies (SFGs) at $z \sim 1-3$ drive powerful gas outflows (e.g., Shapley et al. 2003; Weiner et al. 2009; Kornei et al. 2012; Erb et al. 2012). As part of our "SINS/zC-SINF" imaging spectroscopic survey with SINFONI of ~100 SFGs at $z \sim 2$ (Förster Schreiber et al. 2009; Mancini et al. 2011; hereafter FS09 and M11), we also detected such galactic winds through broad components in the rest-optical H$\alpha$ and [NII]$\lambda\lambda$6548,6584 emission line profiles (Shapiro et al. 2009; Genzel et al. 2011; Newman et al. 2012a, b). These observations allowed us to constrain the origin of the winds within galaxies, and the spatial extent of the outflowing gas essential to derive mass outflow rates. In the 27 best-resolved non-AGN galaxies analyzed in our previous work (Genzel et al. 2011; Newman et al. 2012a, b), the outflows are spatially extended over much of the galaxies, have typical velocity FWHMs of 400–500 km s$^{-1}$ and warm ionized gas mass outflow rates of $\dot{M} \sim 1-8$ times the star formation rates (SFRs), and in several cases originate from luminous kpc-sized star-forming "clumps" that are ubiquitous in $z > 1$ SFGs. These outflows are thus likely driven by supernova explosions, stellar winds, and radiation pressure from hot massive stars (e.g., Hopkins et al. 2012; Genel et al. 2012).

In this paper, we report the detection of a yet broader emission component (typical FWHM ~ 1500 km s$^{-1}$) in the H$\alpha$, [NII], and [SII] lines, originating in the inner few kpc of all of the most massive SINS/zC-SINF galaxies, all of which are large disks with evidence of a significant stellar bulge. This discovery was made possible by our most recent deep high-resolution adaptive optics (AO) assisted SINFONI observations, focussed on $z \sim 2$ SFGs at stellar masses near and above $M_* \sim 10^{11}$ M$_\odot$. The spectral and spatial properties of this "broad nuclear emission" component[1] are highly suggestive of a Type 2 AGN-driven outflow. So far, spatially-resolved observations of warm ionized gas outflows powered by AGN at high redshift exist only for a dozen or so of very luminous but rare radio-detected galaxies, QSOs, and AGN-hosting submillimeter galaxies (Nesvadba et al. 2011; Harrison et al. 2012; Cano Díaz et al. 2012). Our findings provide new constraints on the properties and prevalence of nuclear outflows plausibly driven by AGN in more "normal" massive SFGs at $z \sim 2$.

The paper is organized as follows. Section 2 describes the galaxy sample and SINFONI data sets. Section 3 presents the analysis and properties of the broad emission component. Section 4 discusses the origin of the broad nuclear emission and implications for the evolution of massive galaxies. Throughout, we adopt a Chabrier (2003) stellar initial mass function and a $\Lambda$CDM cosmology with $H_0 = 70$ km s$^{-1}$ and $\Omega_m = 0.3$.

## 2. OBSERVATIONS AND ANALYSIS

### 2.1. Galaxy Sample and SINFONI Observations

The sample discussed in this paper consists of the eight galaxies at $\log(M_*/M_\odot) \geq 11$ (henceforth simply "massive galaxies") observed and detected in H$\alpha$ line emission in our SINS and zC-SINF surveys of $z \sim 2$ SFGs with SINFONI. As for the full SINS/zC-SINF survey, the massive galaxies were drawn from optical spectroscopic surveys of photometrically-selected candidate $z \sim 1.4 - 2.5$ objects. The galaxies were all selected to have an optical spectroscopic redshift such that their H$\alpha$ emission line would fall in spectral regions of high atmospheric transmission and away from the brightest night sky lines, and an estimated observed integrated H$\alpha$ flux of $\gtrsim 5 \times 10^{-17}$ erg s$^{-1}$ cm$^{-2}$ (either from available long-slit spectroscopy or computed from star formation rate and extinction estimates). The massive sample includes six objects from parent samples selected based on $K$-band magnitude and optical/near-IR $BzK$ colors (Kong et al. 2006; M11; S. Lilly et al. in preparation), one object from a $K$–band limited survey (Cimatti et al. 2002; Daddi et al. 2004), and another object from an optical $U_nGR$ color-selected survey (Steidel et al. 2004; Erb et al. 2006). Six galaxies lie at $2.2 < z < 2.4$ and two at $1.5 < z < 1.7$. The combination of color and/or magnitude selection, mandatory optical spectroscopic redshift, and minimum H$\alpha$ flux implies that the galaxies probe the bulk of the massive z ~ 2 SFG population (with some biases discussed below). Table 1 lists the galaxies along with their main properties. Complete details about the target selection, the observations and data reduction, and the derivation of global stellar properties from the galaxies' broad-band optical, near- to mid/far-infrared spectral energy distributions (SEDs) are given by FS09, M11, and Förster Schreiber et al. (2014, in preparation; hereafter FS14); the salient points are summarized below.

---

[1] We call "broad" spectral features that have FWHM velocity widths in the range ~ 400 – 2000 km s$^{-1}$ to distinguish them from narrower FWHM < 300 km s$^{-1}$ line emission from star-forming regions tracing gravitational/turbulent motions. In addition, we use "nuclear regions" to refer to the inner few kpc at the kinematic center of the galaxies. The terminology adopted here should not be confused with the classical "nuclear broad line region" of Type 1 AGN, which refers to the (sub)parsec scale region around the accreting black hole where gas motions are typically characterized by a FWHM > 5000 km s$^{-1}$ (e.g., Osterbrock & Shuder 1982).

The observations were carried out with the near-infrared (near-IR) integral field spectrograph SINFONI (Eisenhauer et al. 2003; Bonnet et al. 2004) at the Very Large Telescope (VLT) of the European Southern Observatory (ESO). The galaxies were all observed in seeing-limited mode at the 125 mas pixel$^{-1}$ scale with typical FWHM resolution of $\approx 0.5''$ (4.2 kpc at z = 2). Six of them were followed up with AO at the 50 mas pixel$^{-1}$ scale either in Natural or Laser Guide Star mode (NGS or LGS) achieving a FWHM resolution of $\approx 0.2''$ (1.7 kpc at z = 2). Depending on the redshift of the targets, the $K$- or $H$-band grating was used to map their H$\alpha$, [NII]$\lambda\lambda$6548,6584, and [SII]$\lambda\lambda$6716,6731 line emission at an effective spectral resolution of $\approx$ 85 and 120 km s$^{-1}$, respectively. For the analysis, we used the AO data sets where available because of their superior angular resolution and signal-to-noise ratio (S/N). However, for the two galaxies with faintest nuclear emission (Deep3a-6004 and Q2343-BX610), we combined their respective AO and no-AO data. While this implies more important beam smearing, the bright clumpy/ring-like structures in the disk of these two galaxies are at sufficiently large radii to not contaminate strongly their nuclear regions spectrum. On-source integrations times were in the range 2h – 10h for the no-AO data sets and 4h – 23h for the AO data sets, with a resulting median of $\approx$ 10h per object. Table 2 summarizes the observational information for the data sets used in this paper.

The SINFONI H$\alpha$ kinematics and distribution show that all the massive galaxies are rotating disks, with large half-light radii ~ 5 – 7 kpc except for one that has a more compact size ~ 2 kpc (Shapiro et al. 2008; FS09; M11; Newman et al. 2013; Genzel et al. 2014; FS14). High-resolution Hubble Space Telescope (HST) broad-band imaging in the $J$ and $H$ bands with the WFC3/IR and NICMOS/NIC2 cameras is available from our follow-up programs for five objects (Förster Schreiber et al. 2011; S. Tacchella et al., in preparation) and from the CANDELS Treasury Survey for one source (Grogin et al. 2011; Koekemoer et al. 2011; Lang et al. 2014). The HST data trace the rest-optical continuum emission from the bulk of the stellar populations. The morphologies and stellar mass surface density maps (derived from the resolved $J$–$H$ colors or 7-band optical/near-IR SEDs) exhibit a disk component and a significant stellar bulge in each galaxy[2]. The mean (and median) of the bulge-to-total stellar mass ratios and of the stellar bulge masses are ~ 0.5 and $\log(M_*/M_\odot)$ ~ 10.8, respectively (S. Tacchella et al., in preparation).

In stellar mass versus star formation rate ($M_*$–SFR), the massive galaxies overlap with the underlying distribution of SFGs in the same z = 1.4 – 2.5 range at $K_{AB}$ < 23 mag and $\log(M_*/M_\odot) \geq 11$ (Figure 1), as do the SINS/zC-SINF full sample and the AO subset (FS14) over the wider $M_*$ range covered. All eight massive galaxies lie within 0.6 dex in log(SFR) of the z = 2 "main sequence" (MS) delineated by the bulk of SFGs in the same redshift and $M_*$ ranges (which has an observed scatter of around 0.3 dex; e.g., Daddi et al. 2007a; Rodighiero et al. 2011; Whitaker et al. 2012). These massive galaxies are thus "normal" in terms of SFR for their stellar masses and redshifts. Nonetheless, they tend to probe the part of the MS with higher specific SFRs and bluer rest-UV/optical colors (FS14). This bias, which applies to the SINS/zC-SINF sample as a whole (FS09; M11), stems primarily from the requirement of a secure optical spectroscopic redshift (implying an inherent rest-UV brightness cut irrespective of the primary photometric selection) and, for the AO subset, the need for an H$\alpha$ detection in the initial 1–2hr seeing-limited observations (further implying that objects with higher surface brightness over at least some part of the source were chosen, e.g., compact sources or extended disks with at least one bright clump). The minimum expected H$\alpha$ flux of $\sim 5 \times 10^{-17}$ erg s$^{-1}$ cm$^{-2}$ imposed when selecting our SINFONI targets probably plays a negligible role at the high-mass end; at z = 2 and for an extinction $A_V = 1 - 2$ mag, this corresponds to an intrinsic SFR $\approx 15 - 30$ M$_\odot$ yr$^{-1}$, a factor of 5 – 10 times or more below the MS at $\log(M_*/M_\odot) \geq 11$.

---

[2] More specifically, the H-band light distributions show a cuspy inner profile or the presence of a red central peak/clump at the kinematic center and four of the six galaxies have best-fit Sersic model indices n > 2.5. More importantly, the stellar mass surface density distributions (which account for M/L variations and better trace the underlying stellar structure) of all six objects exhibit a cuspy inner profile (with best-fit n > 2). Two-component disk + bulge models (with n=1 and n=4, respectively) reproduce much better the stellar mass distributions of the galaxies compared to a single exponential disk model. This echoes the general trends found for massive z ~ 2 galaxies based on high resolution HST imaging of large mass-selected samples by Wuyts et al. (2011b, 2012) and Lang et al. (2014).

In drawing the SINS/zC-SINF targets, we preferentially selected non-AGN sources (although seven confirmed or candidate AGN hosts were included) and none of the selection criteria or observational strategies should introduce an obvious bias towards AGN. Of the eight galaxies at $\log(M_*/M_\odot) > 11$ studied here, four were known or suspected to harbour an AGN at the time of the SINFONI observations based on diagnostic features in their rest-UV spectra, and/or on their X-ray brightness, radio luminosity, or observed mid-IR colors when available (K20-ID5, Deep3a-7144, Deep3a-15504, Q2343-BX610). This larger proportion among the $\log(M_*/M_\odot) \geq 11$ SINS/zC-SINF targets may simply reflect the general increase with galaxy mass of the AGN identification rate at z ~ 2 from rest-UV, mid-IR, X-ray, or radio indicators although several factors complicate comparisons with other studies; we return to this point in Section 4.2. None of the massive SFGs discussed here is a luminous Type 1 AGN with very broad rest-UV/optical emission lines based on the optical spectra obtained for redshift confirmation. Appendix A provides background information from classical AGN diagnostics for the massive galaxies discussed in this paper.

### 2.2. Extraction of Emission Line Spectra and Maps

Visual inspection of the SINFONI data reveals that a broad component is apparent underneath the narrow line emission in each of the eight massive SINS/zC-SINF SFGs. One of them (Deep3a-7144 at *z* = 1.65) is however unsuitable for detailed line profile fitting or inclusion in averaged spectra because of residuals from strong telluric OH lines over a significant fraction of the intervals around the H$\alpha$+[NII] complex. We thus excluded this galaxy from our quantitative analysis.

We extracted emission line spectra from the SINFONI data cubes following procedures as described by Newman et al. (2012b), which optimize the S/N and enhance the contrast between narrow and broad emission components in integrated spectra. Prior to the extraction, the data cubes were smoothed spatially with a 4-pixel wide two-dimensional Gaussian kernel to increase the S/N at the pixel level. Large-scale velocity gradients across the sources were removed by shifting the spectra of individual pixels of a data cube to bring the H$\alpha$ line at the systemic velocity of each galaxy, based on single Gaussian profile fits primarily sensitive to the narrower emission component and tracing the overall gravitationally-driven kinematics. The continuum level was estimated over adjacent line-free spectral intervals and subtracted from the cubes. Nuclear spectra were extracted from the velocity-corrected data cubes in apertures of radius $\approx 0.3''$ (or 2.5 kpc at the redshift of the objects), and disk spectra were extracted outside of the nuclear regions. Co-averaged nuclear and off-nuclear spectra were computed from those of individual galaxies normalized by their peak H$\alpha$ amplitude.

We fitted simultaneously the H$\alpha$+[NII] complex and the [SII] doublet including a narrow and a broad Gaussian component for all five lines, with the following simplifying assumptions. For each component, the relative centroid velocities of the H$\alpha$, [NII], and [SII] lines were fixed and the same width was adopted for all lines. The [NII]$\lambda$6584/H$\alpha$ and [SII]$\lambda$6716/H$\alpha$ ratios were assumed to be identical for the broad and narrow components, and the [NII]$\lambda$6548/$\lambda$6584 ratio was fixed at 0.32. The free parameters were thus the broad and narrow H$\alpha$ centroid velocity and width, the common broad and narrow [NII]$\lambda$6584/H$\alpha$ and [SII]$\lambda$6716/H$\alpha$ flux ratios, and the separate broad and narrow [SII]$\lambda$6716/$\lambda$6731 flux ratios. Such a constrained approach is dictated by the S/N of the spectra of our z ~ 2 galaxies, and the specific assumptions adopted lead to satisfactory fits. All velocity widths quoted in this paper are corrected for the instrumental broadening. The line spread functions (LSFs) of our SINFONI K- and H-band data (as obtained from the profiles of bright isolated night sky lines) show some excess emission in the line wings relative to the best-fit Gaussian. However, these deviations are small compared to the spectral features of interest in our galaxies and, in particular, do not cause the broad emission component analyzed in this paper (see also Genzel et al. 2011, and Appendix B).

For individual pixels, the significant blending and low amplitude of the broad component hampers the reliability of simultaneous multi-component fitting. To extract narrow H$\alpha$ and broad emission flux maps from the data cubes of individual galaxies, we summed the emission over spectral intervals chosen to maximize the respective contributions of the broad and narrow emission (based on the average nuclear regions spectrum) while ensuring sufficient S/N per pixel. For the narrow H$\alpha$ emission maps, integrated in the interval ± 100 km s$^{-1}$ around the reference systemic velocity, the contamination by the broad underlying emission is estimated to be ~ 15% – 20%. For the broad emission maps, computed in the intervals [–1000, –150] and [+150, +700] km s$^{-1}$, the contribution by the narrow emission is ~ 30% – 40% (with respective contributions of ~ 35% – 45% on the blue side including [NII]$\lambda$6548, and of ~ 20% – 30% on the red side). The size estimates based on these maps should thus be representative of the broad emission, which dominates in the adopted velocity intervals.

To independently verify the results based on the spectra and maps extracted for individual galaxies, we followed a second approach by creating a "stacked" data cube of the six massive galaxies observed with AO from which we derived the average H$\alpha$+[NII] broad emission properties. The details of this alternative method are presented in Appendix B. We note that the results from the spectra and maps obtained as described above and from the stacked cube agree well with each other.

## 3. RESULTS

### 3.1. Spectral Properties of the Broad Nuclear Emission

Figure 2 shows the high S/N averaged nuclear spectrum of the H$\alpha$+[NII] emission line complex and the [SII] doublet of the seven massive galaxies whose line profiles are unaffected by telluric OH line emission. From our multiple Gaussian fit to this spectrum, the broad component has a FWHM of $1500 \pm 240$ km s$^{-1}$ in all the lines and contributes about 40% of the total H$\alpha$ and [NII] flux. The [NII]/H$\alpha$ ratio for the broad and narrow components is $0.60 \pm 0.01$, at the upper edge for pure stellar photoionization in metal-rich HII regions and overlapping with the "mixing" sequence of AGN and star formation excitation (e.g., Kewley et al. 2006; 2013).

The fits to the nuclear spectra of individual galaxies yield broad component FWHMs in the range $\sim 600$ to $\sim 1700$ km s$^{-1}$ (median of $\sim 950$ km s$^{-1}$) and contributions to the total H$\alpha$+[NII] flux between 30% and 80% (median of 45%). The [NII]/H$\alpha$ ratios range from 0.5 to 0.75 (median of 0.7). The results are obviously more uncertain for the individual galaxies because of the lower S/N, and the broad nuclear FWHM values tend to be somewhat lower compared to that obtained from the averaged spectrum because the extended wings of the broad low-amplitude component are less well constrained. However, the range in properties derived from the nuclear spectra of the galaxies shows that the results based on the averaged nuclear spectrum are not dominated by a single galaxy, confirming the visual assessment (Section 2.2; see also Appendix B).

Figure 2 also shows the averaged off-nuclear spectrum of the same seven massive galaxies. While the latter spectrum does exhibit a broad component, it is weaker and has a FWHM of $485 \pm 30$ km s$^{-1}$, more typical of the star formation-driven outflows in the lower-mass SINS/zC-SINF galaxies (Newman et al. 2012b). The off-nuclear narrow component has an [NII]/H$\alpha$ ratio of $0.35 \pm 0.01$, well within the range spanned by normal HII regions of roughly solar oxygen abundances in local star-forming galaxies (e.g., Kewley et al. 2006; 2013).

### 3.2. Spatial Extent of the Broad Nuclear Emission

Figure 3 shows the spatial distribution of the broad emission in the six $\log(M_*/M_\odot) \geq 11$ galaxies with deep high-resolution SINFONI+AO data. The Figure also shows the distribution of the narrow H$\alpha$ line emission tracing star formation and of the rest-optical stellar continuum emission. The latter maps are from the HST near-IR imaging available for five of the objects, and from the SINFONI data for the sixth galaxy without HST imaging (Deep3a-6397).

The maps reveal that the broad emission is generally strongest and originates primarily in the inner regions of the galaxies, at/near the kinematic center and the location of the stellar bulge identified from the continuum images. The one exception is Q2343-BX610, which has the weakest nuclear broad emission and smallest bulge. In that case, moderately broad H$\alpha$+[NII] emission (with width and [NII]/H$\alpha$ ratio consistent with star formation-driven outflows) and the wings of the very bright narrow H$\alpha$ line from the luminous star-forming complex south of the nucleus is stronger than the broad nuclear emission within the fixed velocity intervals used to extract the broad emission maps.

In four of the six massive galaxies with SINFONI+AO data, the broad emission is spatially resolved in our data and has an intrinsic size of $\approx 2 - 3$ kpc, either in terms of half-light diameter or of spatial offset between the redshifted and blueshifted broad emission. These intrinsic size estimates are derived from the observed sizes taking into account the spatial point spread function (PSF) derived from SINFONI data of stars obtained at regular intervals during the observations of the galaxies (see FS09, FS14, and Appendix B).

A similar intrinsic size is derived based on the broad emission map extracted from the stacked data cube in Appendix B, where smaller spectral intervals were used to reduce the contribution by the narrow emission. This analysis confirms that the source of broad emission in the nuclear regions of the massive AO targets is on average spatially extended.

### 3.3. Trends with Galaxy Stellar Mass

The presence of a nuclear emission line component with spectral and spatial properties as derived above appears to be strongly mass dependent. Figure 4 compares the emission line properties from the averaged nuclear and off-nuclear spectra of the massive galaxies and of the lower mass SINS/zC-SINF galaxies observed with AO or, in a few cases, sufficiently well resolved in the no-AO data (see Newman et al. 2012b). Only galaxies for which the entire H$\alpha$+[NII] region is not significantly affected by residuals from telluric OH lines are included, and the stellar mass bins contain 5 – 10 SFGs each.

Figure 4 shows that the broad nuclear line width (and the narrow component [NII]/Hα ratio) increases rapidly near $M_* \sim 10^{11}$ $M_\odot$. Below this mass, the nuclear and off-nuclear broad component properties are essentially indistinguishable in our SINS/zC-SINF spectra and are both consistent with star formation-driven winds, as discussed by Newman et al. (2012b). Given the measurement uncertainties, the significance of the upturn in nuclear broad emission FWHM at the high-mass end is 3.4σ; for the highest mass bin, the difference in broad FWHM between nuclear and off-nuclear regions is significant at the 4.2σ level.

The trend of increasing broad component width and narrow emission [NII]/Hα ratio in the nuclear regions above $\log(M_*/M_\odot) \sim 11$ is unlikely to be driven by sensitivity. The co-averaged data represent ~ 30 – 85 hours of integration time per mass bin, and the quality of all resulting nuclear and off-nuclear spectra is outstanding. Beam smearing could play a role in the following sense. While the spectra rely on AO-assisted SINFONI data with essentially identical resolution for all but a few objects, the lower-mass galaxies tend to be smaller. The nuclear apertures (with typical radii of 0.2″ – 0.3″) may thus include a higher contribution from the disk regions, which could dilute a nuclear broad and high-excitation signature. To test this possibility would require sub-kpc resolution data given the decrease by a factor of ~ 2 in average Hα half-light radius from $M_* \sim 10^{11}$ to $10^{10}$ $M_\odot$ among our SINS/zC-SINF SFGs.

## 4. DISCUSSION

### 4.1. The Origin of the Nuclear Broad Emission

A straightforward interpretation of the broad nuclear emission and the high [NII]/Hα ratios in our massive galaxies is that they are caused by an AGN. Indeed, the z > 1 AGN detection rate in X-ray, optical, and infrared observations increases from a few percent at $\log(M_*/M_\odot)$ ~ 10 – 10.5 up to ~ 30% at $\log(M_*/M_\odot) > 11$ (e.g., Reddy et al. 2005; Papovich et al. 2006; Daddi et al. 2007b; Brusa et al. 2009; Hainline et al. 2012; Bongiorno et al. 2012). In addition, recent high-redshift studies have revealed that the AGN host population is drawn from normal MS SFGs (e.g., Mullaney et al. 2012; Rosario et al. 2012, 2013). An AGN is known to be present in half of our massive SINS/zC-SINF galaxies from independent diagnostics. Arguably, the broad nuclear emission in the other massive galaxies could also be related to an AGN that has not yet been identified because of the lack of adequate observational diagnostics for these objects or because the AGN is less luminous and/or more obscured (see Appendix A and Section 4.2).

A key result of our observations is that the broad emission component in our massive galaxies does not trace a classical broad line region, where very dense gas clouds ($n_H \sim 10^{10}$ cm$^{-3}$) orbit around the central black hole on (sub-)parsec scales (e.g., Osterbrock & Shuder 1982). Firstly, the broad emission is spatially resolved in four of our six massive SFGs with SINFONI+AO data, with intrinsic size of ≈ 2–3 kpc. Given this size and the estimated stellar bulge masses, we exclude that the nuclear broad line widths derived for the galaxies are predominantly caused by gravitationally bound orbital motions since the inferred effective circular velocities of the bulges do not exceed ~ 300 km s$^{-1}$, even allowing for an additional 50% gas contribution to the mass. Secondly, the broad emission is clearly detected in the forbidden [NII] and [SII] lines, which implies hydrogen densities of only ~$10^2$ cm$^{-3}$ or less[3]. Therefore, the broad emission line gas cannot all be bound and must be in an outflow, in agreement with an earlier proposal for one of our massive SINS/zC-SINF galaxies (K20-ID5; van Dokkum et al. 2005).

In many respects (outflow velocity, spectral properties), the nuclear outflows in our massive galaxies appear to be scaled-up versions of the nuclear outflow in the nearby archetypal Seyfert 2 galaxy NGC 1068 (Cecil et al. 1990). In more recent integral field spectroscopic studies, Westmoquette et al. (2012) and Rupke & Veilleux (2013) have mapped the gas kinematics and outflow properties from Hα, [NII], and [SII] emission in nearby ultraluminous infrared galaxies (ULIRGs). These authors found that fast (FWHM ≳ 1000 km s$^{-1}$) spatially extended ionized gas outflows are powered by (obscured) AGN. Optical and far-IR observations of atomic (Na I D) and molecular (OH) absorption features in local ULIRGs also showed that AGN drive neutral and molecular gas outflows with highest observed velocities (Sturm et al. 2011; Rupke & Veilleux 2013; Veilleux et al. 2013). Unlike the z ~ 0 ULIRGs in these studies, our massive galaxies are all disks with no evidence of major merging and are not "outliers" with respect to the z ~ 2 MS of SFGs. However, the similarity in spectral and spatial properties of the nuclear broad emission lends further support to our AGN-driven outflows interpretation.

An alternative possibility is that the broad emission and elevated [NII]/Hα ratios in the nuclear regions of our massive galaxies could be due to extremely powerful star formation feedback and shock excitation. The AGN-driven outflow scenario is arguably most compelling for K20-ID5 and Deep3a-15504 (e.g., from their [NII]/Hα and [OIII]/Hβ ratios; see Appendix B and Newman et al. 2014) but these objects are not the most extreme in terms of broad emission FWHM or [NII]/Hα ratio among our sample. Large outflow velocities of up to ~ 1000 – 2000 km s$^{-1}$ have been measured from optical absorption line studies for several local "Lyman-break analogs" and compact post-starburst systems at z ~ 0.6 (Tremonti et al. 2007; Heckman et al. 2011; Diamond-Stanic et al. 2012). The high wind speeds in these systems have been ascribed to extremely high SFR surface densities up to ~ 1000 M$_\odot$ yr$^{-1}$ kpc$^{-2}$, although some contribution from an obscured AGN could not be definitely ruled out. For our massive galaxies, however, the narrow Hα emission in the nuclear regions implies much lower SFR surface densities of ~ 1 – 15 M$_\odot$ yr$^{-1}$ kpc$^{-2}$ (corrected for extinction). These values lie within the range spanned by the brightest off-nuclear star-forming clumps among our SINS/zC-SINF sample and for which the broad emission has a FWHM ~ 400–500 km s$^{-1}$ (Genzel et al. 2011; Newman et al. 2012b). Along with the other lines of evidence above, we thus favor the AGN-driven outflow interpretation for the broad nuclear emission detected in our massive galaxies.

---

[3] The [SII]λ6716/λ6731 broad line ratio of 1.0 ± 0.35 for the average nuclear regions spectrum has a 1-σ upper bound reaching into the low density regime, so that the hydrogen gas density could be < 100 cm$^{-3}$.

*4.2. Frequency of the Nuclear Outflow Phenomenon*

Inspection of the Hα+[NII] spectra for other z ~ 2 log(M$_*$/M$_\odot$) ≥ 11 galaxies in the literature (Erb et al. 2006; Kriek et al. 2007; Swinbank et al. 2012) and in our survey of SFGs with the LUCI multi-object spectrograph at the Large Binocular Telescope (J. D. Kurk, et al., in preparation) supports the prevalence of a broad high-excitation emission component as seen in the nuclear regions of our massive SINS/zC-SINF galaxies. The 16 massive SFGs from these studies are also plotted in Figure 1. While most of them lack high spatial resolution information, the available spectra allow us to identify those that exhibit the combination of strong enough broad Hα+[NII] emission and high [NII]/Hα ratio in their source-integrated spectrum. In the combined sample of 24 SFGs at log(M$_*$/M$_\odot$) ≥ 11, 17 (70%) exhibit the broad component and high excitation signature. This fraction could conceivably be higher since emission from star formation in the outer disk regions could outshine the signature of a fast, high excitation nuclear outflow in the integrated spectra.

The fraction of massive SFGs exhibiting the nuclear AGN-driven outflow signature may appear surprisingly high, given that only half of the broad nuclear emission SFGs in Figure 1 are known to harbour an AGN based on diagnostics relying on rest-UV spectral features, X-ray or radio detection, observed mid-IR colors, and/or rest-optical pairs of emission line ratios such as [OIII]λ5007/Hβ versus [NII]λ6584/Hα. Taken at face value, even this 50% fraction of confirmed AGN among our massive SINS/zC-SINF galaxies is higher than reported for other z ~ 2 galaxy samples at similarly high masses, which are typically in the range ~ 10% – 30% (e.g., Reddy et al. 2005; Papovich et al. 2006; Brusa et al. 2009; Hainline et al. 2012; Bongiorno et al. 2012). On the other hand, it is the same fraction as for the log(M$_*$/M$_\odot$) ≥ 11 galaxies in the z ~ 2 mass-selected sample of Kriek et al. (2007) and comparable to that of mid-IR excess Compton-thick AGN candidates at log(M$_*$/M$_\odot$) ~ 11 reported by Daddi et al. (2007b). Clearly, our sample is very small and the biases resulting from our selection and observing strategy (actively star-forming galaxies with fairly blue rest-optical colors, higher fraction of known AGN observed at log(M$_*$/M$_\odot$) ≥ 11, Hα detection) complicate comparisons with other studies. Furthermore, the different fields from which our SINS/zC-SINF targets were drawn do not have uniform multi-wavelength coverage and/or depth so it is not possible to ascertain reliably the AGN identification rate between different indicators (see Appendix A).

It is also important to bear in mind that the efficacy of AGN selection criteria depends on the intrinsic AGN luminosity, on AGN obscuration, and on whether the emission is host-dominated (e.g., Gilli et al. 2007; Hao et al. 2010, 2011; Hainline et al. 2012; Donley et al. 2012; Mullaney et al. 2012; Aird et al. 2012). A substantial range in these properties could plausibly account for partial cross-identification of AGN between different diagnostics. Also, the nebular line emission excited by an AGN and/or associated with an AGN-driven outflow originates from more extended regions than the X-ray emission, and so is much less affected by the small-scale obscuration and rapid variability of the AGN (e.g., Netzer et al. 2006; Juneau et al. 2011, 2013). One might thus expect a higher occurrence of kpc-sized AGN-driven outflows compared to the fraction of X-ray identified AGN.

An additional consideration is the recombination timescale that, for a hydrogen or electron density n$_e$ ≲ 100 cm$^{-3}$ and assuming an electron temperature of ~ 10$^4$ K for the broad nuclear outflow emission, is ≳ 1000 yr, broadly comparable to the light travel time up to a distance of 1 – 1.5 kpc, which would tend to average out variations of the AGN from episodic accretion on shorter timescales. As we note in Appendix A, shocks could possibly also play a role in the excitation of the observed broad emission for our massive galaxies. Recent simulations (Zubovas & King 2012; Gabor & Bournaud 2014) indicate that AGN-driven outflows may be detected even when the AGN is "off" because the super-heated gas around the AGN expands in a shock propagating out of the host galaxy, maintaining high velocities, over timescales that can be longer than the high accretion rate events onto the black hole. For a representative velocity of 1500 km s$^{-1}$, the outflowing material would take ~ 1 Myr to propagate to a distance of 1.5 kpc. While admittedly uncertain and simplistic, these considerations may suggest that a combination of AGN and shock excitation could maintain the warm outflowing ionized gas highly excited (compared to pure stellar photoionization) and result in a high frequency of the nuclear outflow phenomenon as inferred here.

Beam smearing (and sensitivity) further influence the ability to identify an AGN from near-IR spectroscopic data. For K20-ID5 and Deep3a-7144, only seeing-limited SINFONI data are available but the integrated ratios and line profiles for H$\alpha$ and [NII] suffice to reveal the AGN and associated outflow signature in these objects. Inspection of the seeing-limited SINFONI data for the other galaxies shows that only one of them (zC400528) would be clearly (and individually) identified as hosting an AGN and exhibiting a nuclear outflow signature at no-AO resolution. For all other sources, the [NII]/H$\alpha$ line ratio reaches at most $\approx$ 0.5 in the inner seeing-limited resolution element (FWHM $\approx$ 4 kpc) and the broad emission line component is less prominent, especially for the objects with shortest integration times (FS14). Our study emphasizes the importance of deep and high-resolution observations in identifying and studying the properties and power source of outflows at high redshift.

### 4.3. Implications for Massive Galaxy Evolution

Given the sizes and velocity widths of the broad nuclear emission in our massive galaxies, the implied mass outflow rates in warm ionized gas as traced by our observations are substantial. To estimate the outflow rates, we followed the approach of Genzel et al. (2011) and Newman et al. (2012b) and considered a simple model with a constant wind speed and a constant local volume density of ~100 cm$^{-3}$ (motivated by the [SII]$\lambda$6716/$\lambda$6731 broad line ratio of 1.0 ± 0.35 from the averaged nuclear spectrum). The median outflow rate is $\dot{M}_{out}$ ~ 60 M$_\odot$ yr$^{-1}$ for the seven massive SINS/zC-SINF SFGs with best quality spectra (with a range from 20 to 1,100 M$_\odot$ yr$^{-1}$, all corrected for dust extinction). From the narrow H$\alpha$ emission in the nuclear regions of the galaxies the median intrinsic SFR is ~ 45 M$_\odot$ yr$^{-1}$ (with range of 15 – 130 M$_\odot$ yr$^{-1}$), and the median mass loading factor $\dot{M}_{out}$/SFR is ~ 3 (with range of ~ 0.5 – 15).

The large fraction of broad emission SFGs at high stellar masses implies a high duty cycle of the phenomenon. For roughly constant rates, the nuclear outflows could thus remove gas from the inner regions of the galaxies faster than would be consumed by star formation. Alternatively, for a typical specific star formation rate sSFR ~ 1 Gyr$^{-1}$, the stellar mass doubling time is 0.7 Gyr and an outflow with ~ 60 M$_\odot$yr$^{-1}$ would expel ~ 4 × 10$^{10}$ M$_\odot$ during that time, which is comparable to the stellar bulge mass of our massive galaxies. Even with 50% duty cycle, the total gas mass outflow rates must be very large given that we do not account for the possibly dominant neutral and molecular components (e.g., Rupke & Veilleux 2013; Contursi et al. 2013). While these are crude order-of-magnitude estimates, they do suggest that the nuclear outflow phenomenon in log(M$_*$/M$_\odot$) $\geq$ 11 SFGs at z ~ 2 is playing a significant role in removing gas from the inner regions of massive galaxies.

Recent theoretical work and observations have led to the realization that in the very gas-rich environments of early Universe galaxies, gravitational torques induced by giant star-forming clumps and star formation feedback create large turbulent motions and drive gas and clumps rapidly inward by viscous and dynamical friction (e.g., Genzel et al. 2008, 2011; Dekel et al. 2009, 2013; Bournaud et al. 2011; Ceverino et al. 2012). If so, this process may lead to rapid bulge build-up and concomitant black hole growth. Strongly accreting black holes are in principle capable of converting very efficiently gravitational energy into radiation and winds, hence may be very important in driving gas out of the central regions of their host galaxies. Fast (~ 1000 – 2000 km s$^{-1}$), energetic, kpc-scale warm ionized gas outflows have been recently reported from spatially-resolved rest-optical emission line observations in some luminous z ~ 1.5 – 3.5 systems associated with powerful QSOs (Nesvadba et al. 2011; Harrison et al. 2012; Cano Díaz et al. 2012). Our new observations provide empirical evidence suggesting that efficient and high duty cycle AGN-driven feedback via nuclear outflows may be widespread among the more general population of massive z ~ 2 SFGs.

It is tempting to establish a connection between our findings and the quenching of star formation in massive z ~ 2 galaxies. From a dynamical analysis based on the kinematics and star formation properties as traced by the star formation-dominated narrow H$\alpha$ emission component, Genzel et al. (2014) found that in several of our massive galaxies, the Toomre parameter rises well above the critical value of Q ~ 1 in the inner few kpc where the star formation activity also appears to be suppressed, which would be consistent with central gravitational quenching (e.g., Martig et al. 2009). Recent high-resolution HST near-IR observations of the H$\alpha$ and stellar light distributions at 0.7 < z < 1.5 further support such an "inside-out" quenching process at play in high-mass galaxies (Wuyts et al. 2013; Nelson et al. 2013). If neither this gravitational quenching nor the nuclear outflows discussed in this paper is likely to lead by itself to a long-term galaxy-wide shutdown of star formation, these observations may indicate how both processes could contribute to quench and maintain quiescence in the central regions of massive galaxies.

## 5. SUMMARY

We have uncovered signs of AGN-driven outflows in all of the most massive ($\log(M_*/M_\odot) \geq 11$) $z \sim 2$ SFGs observed with SINFONI as part of our SINS/zC-SINF survey, all of which are disks and host a significant stellar bulge. In contrast to the other SINS/zC-SINF galaxies, which extend to $\sim 20$ times lower stellar masses and exhibit signatures of star formation-driven outflows, the broad component in the massive disks is resolved but more centrally concentrated and, in the inner few kpc, has an elevated [NII]/H$\alpha$ ratio of $0.5 - 0.75$ and larger velocity FWHM up to $\sim 1500$ km s$^{-1}$. This broad emission is detected in H$\alpha$ and forbidden [NII] and [SII] emission and, together with the inferred intrinsic size of $2 - 3$ kpc, is consistent with an outflow driven by an obscured AGN, the presence of which is confirmed from independent diagnostics in half of the galaxies. Examination of published near-IR spectra of other $\log(M_*/M_\odot) \geq 11$ SFGs provides supporting evidence for the prevalence of such AGN-driven nuclear outflows among massive $z \sim 2$ SFGs.

Our results bring new empirical constraints on AGN feedback in "normal" massive main-sequence SFGs at $z \sim 2$, extending previous studies focussed on more extreme systems associated with luminous QSOs. The inferred ionized gas mass outflow rates and mass loading factors for our massive galaxies are substantial and the implied duty cycles for the nuclear outflow phenomenon are high. Our findings suggest that such outflows could play an important role in galaxy evolution, specifically by efficiently clearing the inner regions of massive galaxies from gas.

Both the identification and the study of nuclear outflows in our massive SINS/zC-SINF galaxies benefitted from the high S/N and kpc-scale resolution of our deep SINFONI AO-assisted data. Sensitive and high-resolution near-IR integral field spectroscopy is essential to detect the broad, low-amplitude outflow signature, constrain its spatial extent, and assess differences between nuclear and outer disk regions. Such observations of larger and more complete samples along with high-quality multi-wavelength data will be important to confirm the frequency of the nuclear outflow phenomenon among massive high-redshift SFGs, establish more clearly the association with AGN activity, and constrain more robustly the role of such outflows in early galaxy evolution.

We thank the ESO Paranal staff for their helpful support with the SINFONI observations for this work. We also thank P. van Dokkum, D. Rupke, M. Westmoquette, and A. Cimatti for fruitful discussions, and M. Brightman for further information on the X-ray properties of our targets. We thank the referee for very constructive and insightful comments, which have improved the manuscript. C. M. and A. R. acknowledge support through a PRIN 2010 grant of the Italian National Institute of Astrophysics (INAF).

## APPENDIX A

### EVIDENCE FOR AGN ACTIVITY FROM CLASSICAL DIAGNOSTICS

The massive SINS/zC-SINF galaxies discussed in this paper were drawn from different fields with different multi-wavelength coverage and different depth among the data sets. As a result, the indicators of AGN activity available for our sample form a heterogeneous set. The information for our massive galaxies is described below and summarized in Table A1.

*Rest-UV Spectral Features*

The galaxies all have optical spectra taken for redshift confirmation. Signatures of Type 2 AGN activity are detected in two of them: Deep3a-7144 and Deep3a-15504 (CIV$\lambda$1549, HeII$\lambda$1640, CIII]$\lambda$1909 for both, and also Ly$\alpha$ for the latter). Although HeII$\lambda$1640 and CIII]$\lambda$1909 are detected in the 1150Å – 2000Å spectrum of K20-ID5, the non-detection of other high(er) ionization lines could be consistent with the rest-UV line emission being dominated by photoionization by hot massive stars (Daddi et al. 2004; Shapley et al. 2003).

*X-ray and Radio Emission*

X-ray and radio observations are available for the Chandra Deep Field South (CDFS) and COSMOS fields encompassing three of our massive galaxies. In CDFS, K20-ID5 is detected in both hard (2–10 keV) and soft (0.5–2 keV) X-ray bands in the deep 4 Ms Chandra data (Xue et al. 2011), with properties indicating a Compton-thick AGN of intrinsic rest-frame hard band $\log(L_X / \text{erg s}^{-1}) \approx 43.6$ (Brightman & Ueda 2012)[4]. This object is also detected in radio emission (Daddi et al. 2004; Miller et al. 2008; Bonzini et al. 2012) and its VLA 1.4GHz flux density of 110$\mu$Jy would imply a SFR $\approx 1095$ M$_\odot$ yr$^{-1}$ (adopting the conversion of Ivison et al. [2010a,b] scaled to the Chabrier IMF), significantly higher than its $SFR_{UV+IR} \approx 385$ $M_{\odot}$ $yr^{-1}$ and thus indicative of an AGN contribution to the radio emission.

---

[4] The non-detection of K20-ID5 in hard X-ray in earlier, shallower 1Ms Chandra observations along with its low soft X-ray to optical luminosity ratio and the lack of unambiguous rest-UV AGN signatures had initially led to the conclusion that this object is dominated by vigorous star formation activity (Daddi et al. 2004; see also van Dokkum et al. 2005).

The X-ray and radio data in COSMOS are shallower than in CDFS as a result of the trade-off between area and sensitivity. None of the two zC-SINF sources is detected in the deepest X-ray observations with Chandra (Elvis et al. 2009); their flux upper limits imply $\log(L_X / erg\ s^{-1}) <$ 44 assuming a photon index $\Gamma = 1.9$ and intrinsic line-of-sight absorption column density of $N_H = 10^{22}$ $cm^{-2}$ (these limits could be substantially higher in the Compton-thick case). In the COSMOS VLA radio data (Schinnerer et al. 2010), only zC400528 is detected and its 1.4GHz flux density of 67μJy would imply a SFR $\approx 790$ $M_{\odot}$ $yr^{-1}$, exceeding its $SFR_{UV+IR} \approx 550$ $M_{\odot}$ $yr^{-1}$. Given the uncertainties in both estimates, the difference for this object is not very significant but does not rule out AGN activity.

*Observed Mid-Infrared Colors*

Observed mid-IR colors measured with the Spitzer/IRAC instrument have been argued to offer a powerful technique to select luminous AGN through the emission of dust heated to ~ 1000 – 1500 K in their vicinity, including heavily obscured AGN that are missed in the X-ray (Hao et al. 2010, 2011). IRAC imaging is available for the GOODS-South and COSMOS fields in all four bands (3.6, 4.5, 5.8, and 8.0 μm), and in some subsets of the bands for the survey fields from which our $U_nGR$ color-selected "BX" targets were taken. K20-ID5 satisfies the AGN criteria in the $\log(S_{5.8}/S_{3.6})$ versus $\log(S_{8.0}/S_{4.5})$ diagram of Lacy et al. (2007) as recently revised for high-redshift samples by Donley et al. (2012) as well as those in [5.8]–[8.0] and [3.6]–[4.5] color space of Stern et al. (2005). The two zC-SINF sources lie > 2σ away from the AGN "wedge" in the $\log(S_{5.8}/S_{3.6})$ – $\log(S_{8.0}/S_{4.5})$ plane, and so does zC400569 in the [5.8]–[8.0] versus [3.6]–[4.5] color diagram while zC400528 lies essentially on the boundary. Q2343-BX610, observed in the IRAC 4.5 and 8.0 μm bands, has $\log(S_{8.0}/S_{4.5})$ within 1σ of the AGN boundary of Donley et al. (2012).

*Rest-Optical Line Ratios*

Ideally, the presence of an obscured AGN in our massive galaxies would be identified from a pair of rest-optical emission line ratios allowing us to disentangle between different excitation mechanisms, such as the classical diagnostic diagram combining [NII]λ6584/Hα and [OIII]λ5007/Hβ (Baldwin et al. 1981; Veilleux & Osterbrock 1987; Kewley et al. 2006; Sharp & Bland-Hawthorn 2010). In practice, this method is observationally challenging at z ~ 2 because it requires that both [OIII] and Hβ also fall in clear spectral regions away from night sky lines, and the wavelength separation and weakness of one or both of these lines relative to Hα generally necessitate long integrations in a different instrument setup. An additional complication is that the widely used boundaries between OB star photoionization and AGN-dominated excitation based on local galaxies may not be appropriate for high redshift galaxies because of different ISM conditions (e.g., Liu et al. 2008; Kewley et al. 2013).

For four of the eight massive SINS/zC-SINF galaxies, we obtained additional seeing-limited SINFONI data in the *H* band to map the [OIII]+Hβ emission and complement the *K* band observations of Hα+[NII] (Förster Schreiber et al. 2006; Newman et al. 2014). The source-integrated and nuclear region line ratios of K20-ID5 and Deep3a-15504 fall well within the AGN region of the diagram, whether we consider the criteria based on local galaxies or the proposed versions for z ~ 2 by Kewley et al. (2013). The integrated and nuclear ratios for Q2343-BX610 lie in the so-called "composite region" encompassing the range between the boundaries of local normal SFGs and AGN-dominated systems, and so do the integrated ratios for Deep3a-6004 (the S/N for [OIII] and Hβ in the nuclear regions of this source is too low for reliable measurements); according to the Kewley et al. (2013) models for ISM conditions arguably more appropriate for z ~ 2 galaxies, the ratios for these two objects also lie in the region of overlap between the star-forming abundance sequence and the base of the AGN mixing sequence. If the line excitation for Q2343-BX610 and Deep3a-6004 is more ambiguous, the broad nuclear emission in these two sources is also among the weakest of our massive sample and the measured ratios do not rule out a small AGN contribution. Moreover, beam smearing dilutes the nuclear regions emission more importantly in the seeing-limited data. Higher nuclear [NII]/Hα ratios are generally measured from the SINFONI+AO data including Q2343-BX610 and Deep3a-6004 (FS14), so their actual nuclear line excitation could be higher also in [OIII]/Hβ.

Additional diagnostic ratios from [SII]λ6716+6731/Hα and [OI]λ6300/Hα would be useful to better constrain the nebular excitation mechanism and explore the role of shocks. These lines are weak and their fluxes in the individual galaxies are very uncertain but the average nuclear spectrum yields a broad+narrow [SII]/Hα $\approx 0.3$ and [OI]/Hα $\approx 0.15$, while the average disk spectrum implies [SII]/Hα $\approx 0.4$ ([OI] is essentially undetected in our disk spectrum). These values support the results from [OIII]/Hβ and [NII]/Hα in terms of favoring dominant AGN excitation in the nuclear regions, if the [OIII]/Hβ ratios of the four galaxies discussed above are representative of all seven galaxies included in our average spectra and assuming local galaxies' ISM conditions (e.g., Kewley et al. 2006; Sharp & Bland-Hawthorn 2010). It is however important to keep in mind that more than one mechanism may contribute to the excitation and, in particular, shocks in the outflowing gas together with stellar photoionization could play a role or even mimic AGN-like rest-optical line ratios (e.g., Rich et al. 2010, 2011; Newman et al. 2014). Deeper high-resolution data and/or larger samples will be required to address this issue quantitatively.

## APPENDIX B

## STACKING ANALYSIS OF THE SINFONI+AO DATA OF THE MASSIVE SAMPLE

To strengthen our results on the size estimates and spectral properties of the nuclear broad emission in the massive SINS/zC-SINF galaxies, we complemented our analysis based on the spectra and maps extracted from the individual galaxies' data cubes with one based on "stacking" of the reduced data at the cube level. The main motivation was to achieve sufficient S/N per spatial pixel in order to extract an average high-resolution broad emission map based on narrower velocity intervals, further reducing the narrow emission contamination. Another motivation was to validate our assumption of identical [NII]/H$\alpha$ ratio for the broad and narrow components. For this purpose, we included only the higher resolution data of the six galaxies observed with AO and focussed on the H$\alpha$+[NII] emission.

The velocity-corrected data cubes of the targets were first spatially-registered based on the centroid of the continuum emission, as determined from 3$\sigma$-clipped averages in the line-free intervals [–5000, –3500] km s$^{-1}$ and [+3500, +5000] km s$^{-1}$. This centroid position is well constrained for the massive galaxies, since their rest-optical continuum morphologies in our SINFONI data as well as in the near-IR HST imaging when available exhibit a significant peak coinciding with the dynamical center and attributed to a stellar bulge component (Figure 3; Förster Schreiber et al. 2011; Genzel et al. 2014; S. Tacchella et al., in preparation). The registered velocity-corrected cubes were then continuum subtracted and weight-averaged. The weight was taken as $w = (F_{H\alpha,tot} \times T_{int} \times \sigma_{pix}^2)^{-1}$, where $F_{H\alpha,tot}$ is the total H$\alpha$ flux and $T_{int}$ is the total on-source integration time of each galaxy, and $\sigma_{pix}$ is the rms deviation of every pixel in 3D space from the associated noise cube. The flux and integration time normalization avoids the resulting stack being dominated by the brighter galaxies or by the ones with deepest integration (hence overall highest S/N). The $\sigma_{pix}^2$ term minimizes the impact of noise, in particular in spectral intervals corresponding to bright night sky lines.

The narrow emission map was obtained from the resulting stacked cube using the same velocity interval of $\pm 100$ km s$^{-1}$ around the reference systemic velocity as for the individual galaxies. Taking advantage of the higher S/N per pixel of the stack, narrower intervals were used to create the map of the broad emission, specifically [–470, –200] km s$^{-1}$ on the blue side and [+335, +605] km s$^{-1}$ on the red side, to reduce the contamination from the narrow emission component. The resulting maps are shown in Figure B1. As for the individual galaxies, the broad emission is substantially more centrally concentrated than the narrow emission, and peaks at the center of the "average" galaxy. The size of the broad emission region was estimated from the half-light radius $R_{1/2}$ based on the cumulative curve-of-growth in circular apertures and the HWHM of the azimutally-averaged surface brightness profile, correcting for beam smearing in the stacked data using the co-averaged PSF image of the galaxies. This average PSF was obtained from observations of stars taken for every hour of integration on the galaxies, and thus takes into account the typical narrow core and broad wings of AO PSFs. The radial profiles and curves-of-growth from both the average broad emission and the average PSF are compared in Figure B1. This comparison shows that the average broad emission source is more extended than the PSF, confirming the results from the individual galaxies. The intrinsic $R_{1/2}$ and HWHM imply an average radius of around 1.6 kpc for the central source of broad emission. This size may represent a conservative lower estimate because we did not account for possible differences in isophotal shapes and P.A. of the broad emission among the galaxies when stacking the cubes, so that broad emission in the outer edges of the nuclear regions would tend to be averaged out.

The average nuclear spectrum was extracted from the stacked cube in a circular aperture of radius 0.3″ (2.5 kpc; enclosing $\sim$50% of the broad emission), and is plotted in Figure B1. The average LSF of the data is also plotted for comparison; small excess emission in the low-amplitude wings of the LSFs have a negligible impact on the observed line emission profiles from the galaxies. A multi-component Gaussian fit to the H$\alpha$+[NII] complex was performed, with fixed central wavelengths, narrow and broad line widths, and [NII]$\lambda$6548/[NII]$\lambda$6584 ratio as described in Section 2.2 but the relative intensity of [NII]$\lambda$6584 to H$\alpha$ was allowed to vary between the narrow and broad components. This fit based on the stacked cubes of the six most massive galaxies with AO data yields similar spectral characteristics as derived from the averaged nuclear spectrum of the seven massive galaxies discussed Section 3.1, with small differences attributed to the different averaging method and assumptions on [NII]/H$\alpha$, and to the exclusion of K20-ID5 observed in seeing-limited mode only (which is one of the sources with more prominent nuclear broad emission and higher [NII]/H$\alpha$ ratio). More specifically, the best fit yields an intrinsic velocity FWHM of 1450 km

s$^{-1}$ for the broad component, and the [NII]/H$\alpha$ ratios are 0.83 and 0.53 for the broad and narrow components, respectively. The broad H$\alpha$ emission component contributes about 35% of the total H$\alpha$ emission measured in the inner 5 kpc (diameter) of the stacked AO cube. Based on the best line profile fit, the contribution from narrow emission to the smaller intervals used to make the broad emission map in Figure B1 is about 22%.

We assessed the robustness of the size and line profile properties of the broad emission via jackknifing, constructing a stacked cube and repeating the analysis above excluding each galaxy in turn. Among the jackknife subsamples, the broad emission in the central regions is always found to be spatially extended (intrinsic $R_{1/2}$ in the range 1.7 – 2.1 kpc), and to have an intrinsic velocity FWHM in the range 995 – 1500 km s$^{-1}$. The analysis based on the stacked cube of the six massive galaxies with AO data also indicates [NII]/H$\alpha$ ratios > 0.5 for the broad and narrow components in the nuclear regions. Typically higher ratios are derived for the broad emission, which could reflect an increased contribution by shocks to the broad emission tracing the outflowing gas. We note however that satisfactory fits are also obtained when forcing the broad and narrow [NII]/H$\alpha$ to be equal (giving [NII]/H$\alpha$ ratios > 0.5), as assumed in Section 2.2. The jackknifing further confirms that the observed spectral and spatial properties are not dominated by a single galaxy.

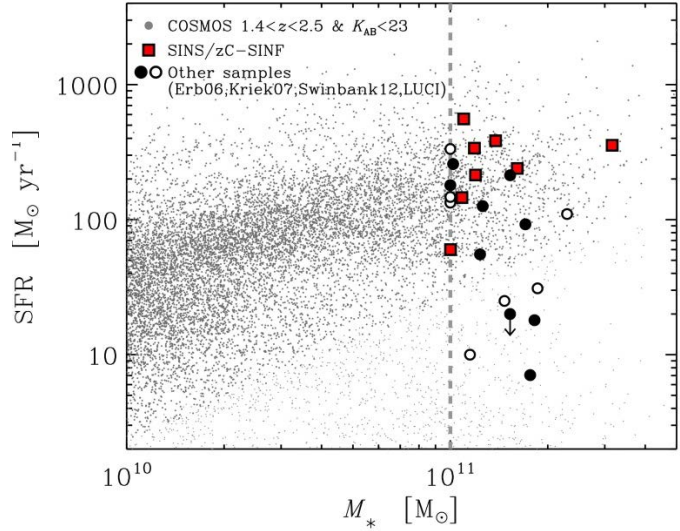

**Figure 1**. Distribution of the z~1.5–2.5 log($M_*/M_\odot$) ≥ 11 galaxies discussed in this paper in the stellar mass versus SFR plane. Filled red squares correspond to the eight most massive galaxies from our SINS/zC-SINF survey at z ~ 2 with SINFONI. All these SFGs exhibit a broad nuclear emission in Hα, [NII], and [SII] lines, and [NII]/Hα flux ratios ≥ 0.5 in the broad and narrow component. Circles indicate massive SFGs from other near-IR spectroscopic studies discussed in Section 4.2 (Erb et al. 2006; Kriek et al. 2007; Swinbank et al. 2012; J. D. Kurk et al., in preparation [LUCI]), for which the available spectra allow distinguishing between galaxies that do (filled black circles) or do not (open black circles) exhibit the same combination of broad emission and high [NII]/Hα ratios. The distribution of the underlying galaxy population at z ~ 1.5 – 2.5 is plotted with small grey dots (from Wuyts et al. 2011b); the "main sequence" of SFGs is apparent as the tight locus containing a majority of all galaxies.

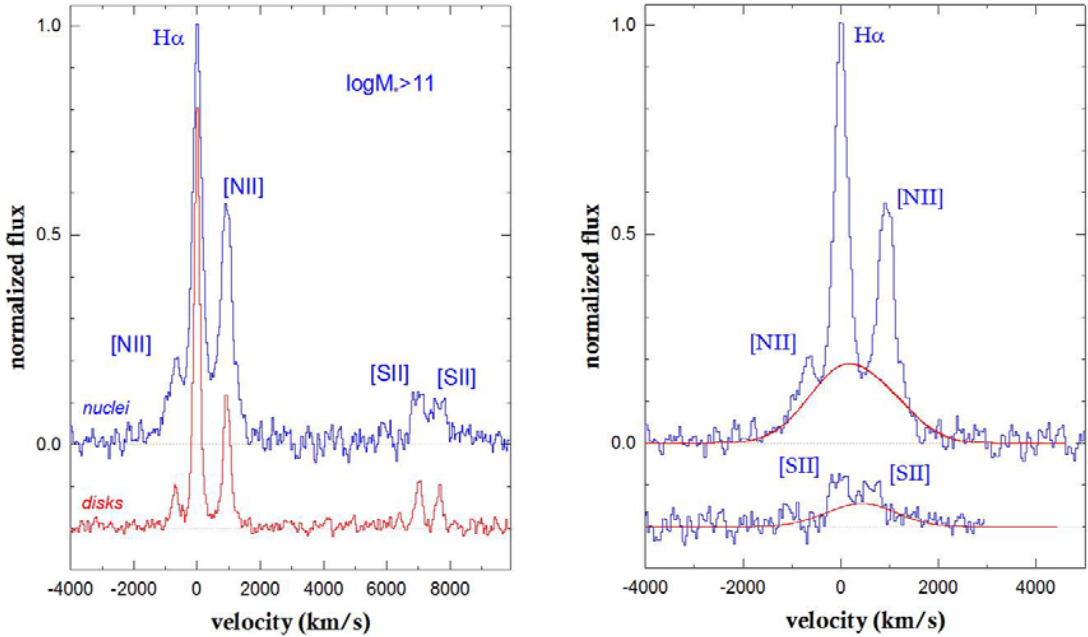

**Figure 2**. Averaged SINFONI spectra covering the Hα, [NII], and [SII] emission lines of the seven SINS/zC-SINF galaxies at log($M_*/M_\odot$) ≥ 11 and best-quality data. *Left*: Averaged spectra of the inner nuclear/bulge (blue) and extra-nuclear/disk regions (red), after removing large-scale velocity gradients from Gaussian fits to the narrow Hα profile in each spatial pixel. *Right*: Zoom-in on the Hα+[NII] complex (top) and the [SII] doublet (bottom) of the nuclear spectrum, and best-fit broad component (red lines). The FWHM of the broad nuclear emission is 1500 ± 240 km/s, and the [NII]λ6584/Hα ratio is 0.60 ± 0.01 for this average spectrum.

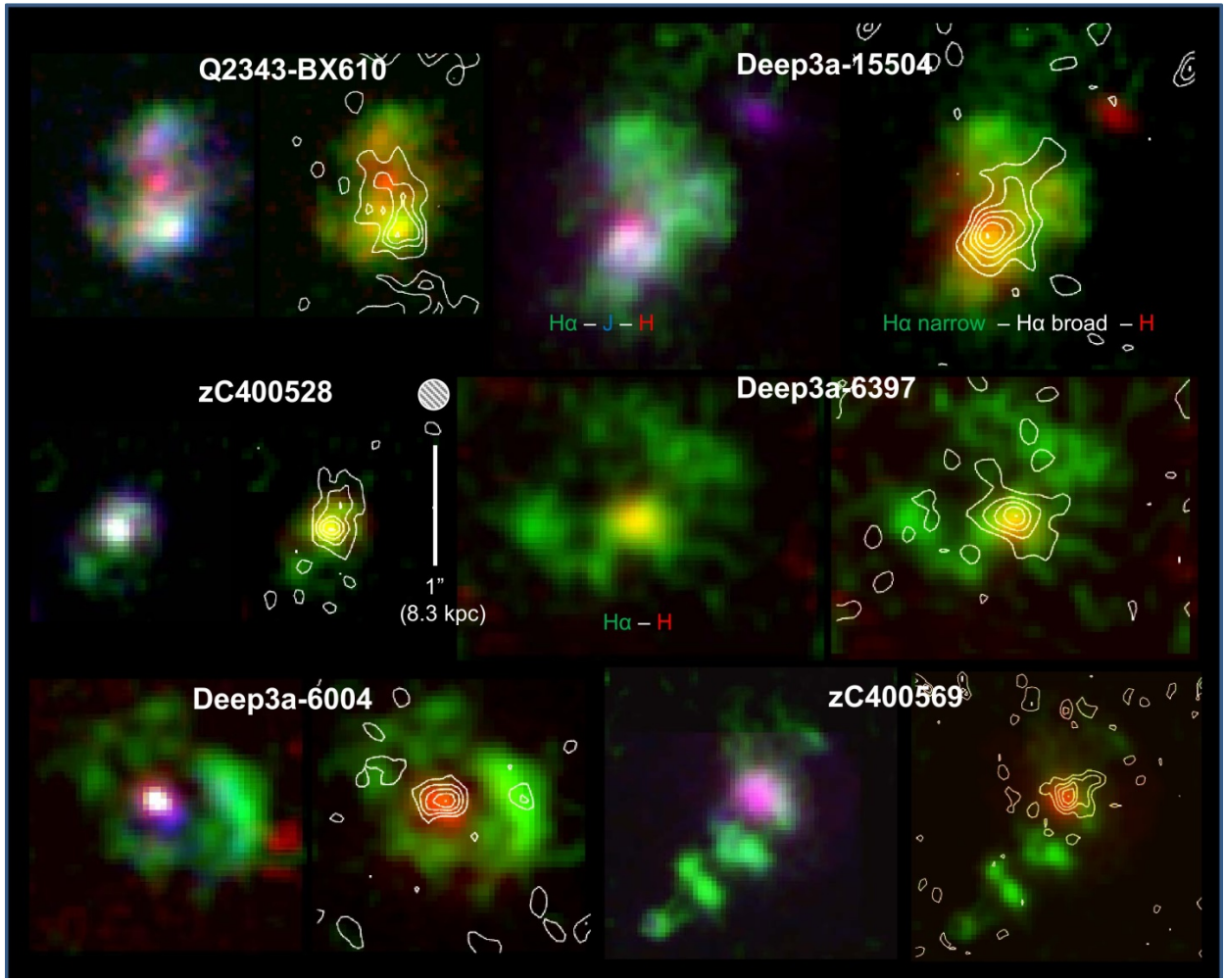

**Figure 3.** Spatial distribution of the broad line emission (white contours) overlaid on maps of the narrow Hα emission (green colours) from the deep high-resolution SINFONI+AO observations obtained for six of the log(M$_*$/M$_\odot$) ≥ 11 SINS/zC-SINF galaxies. The rest-frame ~3500Å and ~4700Å continuum emission maps are also shown (blue and red colours) for the five galaxies with near-IR HST *J* and *H* band imaging with NICMOS/NIC2 or WFC3/IR; for Deep3a-6397, the rest-frame ~6500Å continuum map from the SINFONI+AO data is shown (red colours). The galaxies are shown on the same angular scale (white vertical bar), and all maps have a similar resolution of FWHM ≈ 0.2″–0.3″ (filled circle). All galaxies show central stellar bulges. The strongest broad emission is centered on these bulges except for Q2343-BX610, for which emission from the luminous star-forming complex south of the nucleus is stronger than the broad nuclear emission within the velocity intervals used to extract the broad emission maps (see Section 3.2).

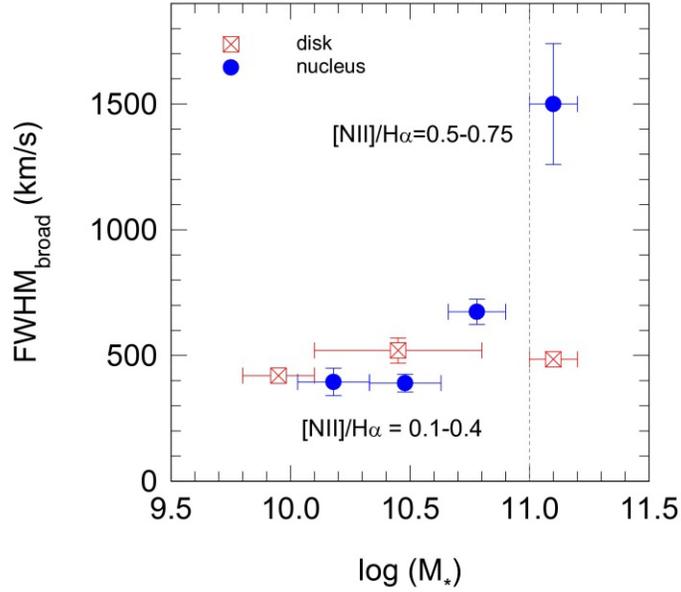

**Figure 4.** Variations of the broad component FWHM and narrow component [NII]/Hα ratio with galaxy stellar mass from the full SINS/zC-SINF sample. The nuclear (blue filled circles) and off-nuclear/disk (red crossed squares) spectra were co-averaged in bins of stellar mass. Each of the bins contains 5–10 SFGs whose spectrum around the Hα+[NII] complex is unaffected by telluric OH line residuals. The broad emission component in the nuclear and off-nuclear regions has similar properties among galaxies below $\log(M_*/M_\odot) \sim 11$. Above this mass, a distinct component appears in the nuclear regions, which has a significantly broader line width indicative of higher outflow velocities and elevated [NII]/Hα ratio consistent with a contribution by an AGN to the nebular line excitation.

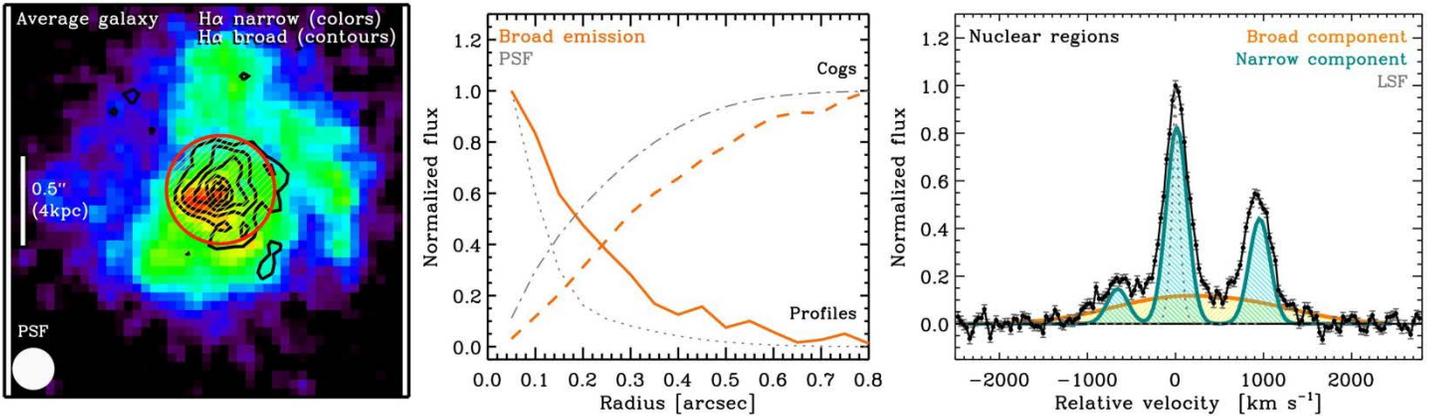

**Figure B1**: Results from the analysis of the co-averaged data cube of the six most massive SINS/zC-SINF galaxies with SINFONI+AO observations. *Left*: Spatial distribution of the broad line emission (contours) and narrow Hα emission (colors). The effective PSF FWHM is shown by the white-filled circle at the bottom left, and the angular scale is indicated by the vertical bar in the middle left of the panel. The spectral intervals used to extract the broad and narrow emission maps are given in the text of Appendix B. The red circle shows the region in which the nuclear spectrum was integrated. Middle: Comparison of the azimutally-averaged surface brightness profile and curve-of-growth of the broad emission from the stacked cube (solid and dashed orange curves) with those of the average PSF (dotted and dash-dotted grey curves), as detailed in the text. The observed broad emission is significantly more extended than the average PSF. Right: Details of the line profile decomposition (as described in text of Appendix B) for the nuclear regions spectrum, showing the observed spectrum (black line), the best-fit broad and narrow components (pale orange and blue), and the average spectral response function (grey dotted line). The average spectral and spatial properties obtained from the stacked cube and those based on spectra and maps extracted from the cubes of individual galaxies agree very well.

**Table 1**
Global Properties and Parent Sample of the SINS/zC-SINF Massive Galaxies

| Galaxy | R.A. (J2000) | Decl. (J2000) | $z_{H\alpha}$ | $\log(M_*/M_\odot)$ | $\mathrm{SFR_{SED}}$ ($M_\odot\,\mathrm{yr}^{-1}$) | $\mathrm{SFR_{UV+IR}}$ ($M_\odot\,\mathrm{yr}^{-1}$) | $\mathrm{SFR_{H\alpha}}$ ($M_\odot\,\mathrm{yr}^{-1}$) | Parent Survey and Selection[a] | References |
|---|---|---|---|---|---|---|---|---|---|
| K20-ID5 | 03:32:31 | −27:46:23 | 2.2243 | 11.1 | 310 | 385 | 235 | K20; *K* | 1,2 |
| zC400528 | 09:59:48 | +01:44:19 | 2.3876 | 11.0 | 150 | 555 | 85 | zCOSMOS-Deep; *K+BzK* | 3,4 |
| zC400569 | 10:01:09 | +01:44:28 | 2.2405 | 11.2 | 240 | 240 | 255 | zCOSMOS-Deep; *K+BzK* | 3,4 |
| Deep3a-6004 | 11:25:04 | −21:45:33 | 2.3867 | 11.5 | 215 | 355 | 570 | Deep3a; *K+BzK* | 5 |
| Deep3a-6397 | 11:25:11 | −21:45:07 | 1.5138 | 11.1 | 565 | 215 | 515 | Deep3a; *K+BzK* | 5 |
| Deep3a-7144 | 11:24:58 | −21:43:57 | 1.6541 | 11.1 | 200 | 340 | 300 | Deep3a; *K+BzK* | 5 |
| Deep3a-15504 | 11:24:16 | −21:39:31 | 2.3826 | 11.0 | 150 | 145 | 215 | Deep3a; *K+BzK* | 5 |
| Q2343-BX610 | 23:46:09 | +12:49:19 | 2.2103 | 11.0 | 60 | ... | 115 | BX/BM; $U_nGR$ | 6,7 |

**Notes.** Main properties of the massive SINS/zC-SINF galaxies studied in this paper. The stellar mass and $\mathrm{SFR_{SED}}$ were derived from stellar evolution synthesis modeling of the observed broad-band optical to near-/mid-IR SEDs following standard procedures, and assuming a Chabrier (2003) stellar initial mass function. The stellar masses correspond to the mass in stars alive and in remnants for the age of the best-fit stellar population model. The $\mathrm{SFR_{UV+IR}}$ estimates for the objects in fields with mid-IR Spitzer/MIPS 24μm and/or Herschel PACS 60–160μm observations were computed from the observed UV and IR luminosities (following Wuyts et al. 2011a). The $\mathrm{SFR_{H\alpha}}$ estimates were calculated from the source-integrated luminosities in the narrow Hα component, corrected for dust extinction assuming that the nebular line emission is 2.3 times more attenuated than the stellar continuum light (Calzetti et al. 2000). Complete details about the target selection for the SINFONI observations and the derivation of their global properties are given by FS09, M11, and FS14.

[a] Parent survey from which the galaxies were drawn and primary photometric criteria used to select the candidate z ~ 2 sources for the optical spectroscopic redshift confirmation (*K*-band magnitude, *BzK* colors, or $U_nGR$ colors).

**References.** The main references for the parent surveys are as follows: (1) Cimatti et al. 2002; (2) Daddi et al. 2004; (3) Lilly et al. 2007; (4) Lilly et al. 2009; (5) Kong et al. 2006; (6) Steidel et al. 2004; (7) Erb et al. 2006.

**Table 2**
Summary of the SINFONI Observations for the Massive SINS/zC-SINF Sample

| Galaxy | Band[a] | Pixel scale (arcsec) | Mode[b] | $T_{int}$[c] (s) | PSF FWHM[d] (arcsec) | ESO Program ID[e] |
|---|---|---|---|---|---|---|
| K20-ID5 | $K$ | 0.125 | No-AO | 9600 | 0.51 | 074.A-9011 |
| zC400528 | $K$ | 0.05 | NGS-AO | 14400 | 0.15 | 183.A-0781 |
| zC400569 | $K$ | 0.05 | NGS-AO | 81000 | 0.17 | 183.A-0781, 091.A-0126 |
| Deep3a-6004 | $K$ | 0.05 | LGS-AO | 19200 | 0.16 | 183.A-0781, 091.A-0126 |
| | | 0.125 | No-AO | 36000 | 0.53 | 076.A-0527, 078.A-0600 |
| Deep3a-6397 | $H$ | 0.05 | LGS-AO | 30600 | 0.19 | 082.A-0396 |
| Deep3a-7144 | $H$ | 0.125 | No-AO | 7200 | 0.54 | 076.A-0527 |
| Deep3a-15504 | $K$ | 0.05 | NGS-AO | 82800 | 0.16 | 076.A-0527, 183.A-0781 |
| Q2343-BX610 | $K$ | 0.05 | LGS AO | 30000 | 0.24 | 183.A-0781, 088.A-0202 |
| | | 0.125 | No-AO | 10800 | 0.39 | 075.A-0466 |

[a] Near-IR band covered by the SINFONI grating used to map the H$\alpha$+[NII]+[SII] line emission of the targets.
[b] Observing mode for the SINFONI data used in this work: "No-AO" for seeing-limited observations, and "NGS-AO" or "LGS-AO" for observations using a Natural Guide Star or a Laser Guide Star for the AO correction.
[c] Total on-source integration time of the observations.
[d] The PSF FWHM reported here represents the angular resolution of the final reduced data of the galaxies, determined from the combined images of a star observed for every hour of integration time; it does not include the effects of the smoothing applied when extracting flux and kinematic maps.
[e] The observations were generally collected over several nights, spread over a few years in some cases, as part of the ESO programs with ID numbers listed. Complete details are given by FS09, M11, and FS14.

**Table A1**
AGN Identification from Complementary Multi-wavelength Diagnostics

| Galaxy | Rest-UV features | Hard band X-ray flux (erg s$^{-1}$ cm$^{-2}$) | Soft band X-ray flux (erg s$^{-1}$ cm$^{-2}$) | 1.4GHz flux density ($\mu$Jy) | Mid-IR colors | Rest-optical line ratios [OIII]/H$\beta$ vs [NII]/H$\alpha$ |
|---|---|---|---|---|---|---|
| K20-ID5 | No | $1.1 \times 10^{-16}$ | $7.6 \times 10^{-17}$ | 110 | Yes | AGN |
| zC400528 | No | $<1.4 \times 10^{-15}$ | $<3.6 \times 10^{-16}$ | 67 | No | ... |
| zC400569 | No | $<1.4 \times 10^{-15}$ | $<3.7 \times 10^{-16}$ | <30 | No | ... |
| Deep3a-6004 | No | ... | ... | ... | ... | Composite region |
| Deep3a-6397 | No | ... | ... | ... | ... | ... |
| Deep3a-7144 | Yes | ... | ... | ... | ... | ... |
| Deep3a-15504 | Yes | ... | ... | ... | ... | AGN |
| Q2343-BX610 | No | ... | ... | ... | Yes | Composite region |

**Notes.** Identification of AGN activity from various diagnostics for the SINS/zC-SINF galaxies at $\log(M_*/M_\odot) \geq 11$. A triple-dot means that no data in the relevant wavelength regime are available. Details of the diagnostics considered and references for the data are described in the text of Appendix A. For four of the eight galaxies (K20-ID5, Deep3a-7144, Deep3a-15504, Q2343-BX610), the presence of a Type 2 AGN is confirmed by one or more of the diagnostics considered. For the other galaxies, a clear identification from these diagnostics is currently hampered by the lack (or shallowness) of the relevant observations, or by a lower luminosity and/or higher obscuration of the AGN.